\documentclass[prb,two column,amsmath]{revtex4}
\input {epsf.sty}
\usepackage{epsfig}
\usepackage{amssymb}
\begin{document}
\title{Anisotropy and Penetration Depth of MgB$_{2}$ from ${}^{11}$B NMR}
\author{Bo Chen$^{1}$, Pratim Sengupta$^{1}$, W. P. Halperin$^{1}$, 
E. E. Sigmund$^{2}$, \mbox{V. F.
Mitrovi{\'c}$^{3}$,} M. H. Lee$^{4}$, K. H. Kang$^4$, B. J. Mean$^4$, J. Y. Kim$^5$, B. K. Cho$^{5}$}
\affiliation{${}^{1}$Department of Physics and Astronomy, 
Northwestern University, Evanston,
Illinois 60208\\
              ${}^{2}$Department of Radiology, New York University, 
New York, New York 10016\\
              ${}^{3}$Department of Physics,  Brown University, 
Providence, Rhode Island, 02912\\
              ${}^{4}$Department of Physics, Konkuk University, Seoul 
143-701, Korea\\
              ${}^{5}$Center for Frontier Materials, Department of 
Materials Science and Engineering,
KJ-IST 500-712, Korea}

\date{Version \today}

\begin{abstract}The ${}^{11}$B NMR spectra in polycrystalline MgB$_2$ 
were measured for several
magnetic fields (\mbox{1.97 T} and 3.15 ~T) as a function of temperature 
from 5 K to 40 K.  The composite
spectra in the superconducting state can be understood in terms of 
anisotropy of the upper critical
field,
$\gamma_{H}$, which is determined to be 5.4 at low temperature.  Using Brandt's
algorithm\cite{Brandt} the full spectrum, including satellites, was 
simulated for the temperature 8
K and a magnetic field of 1.97 T. The penetration depth $\lambda$ was 
determined to be $1,152\pm50$
{\AA}, and the anisotropy of the penetration depth, 
$\gamma_{\lambda}$, was estimated to be close to
one at low temperature. Therefore, our findings establish that there are two different
anisotropies for upper critical field and penetration depth at low temperatures.
\end{abstract}


\maketitle

\vspace{11pt}

The discovery of unusually high superconductive transition 
temperatures of MgB$_2$, a simple
bimetallic compound superconductor\cite{R1}, has attracted 
considerable interest from both theory
and experiment. Reconsideration and extension of BCS theory to 
two-band superconductivity has
successfully accounted for experimental observations\cite{
Choi,Lyard,Rydh,Cubitt,Kogan1,Kogan2}.  Nonetheless, the relation 
between anisotropy of the upper
critical field and the penetration depth is still a controversial 
issue. Generally there are two
points of view. One holds that there exist two different anisotropies at low
temperatures, $\gamma_{H}$ and $\gamma_{\lambda}$, for upper critical 
field and penetration depth
respectively. They have different temperature dependence and merge at 
a common value at
T$_c$\cite{Kogan1,Golubov,Fletcher}. The other perspective is that 
there is only one anisotropy
parameter, and it is field 
dependent\cite{Lyard,Angst1,Zehetmayer}. Moreover, small 
angle
neutron scattering (SANS) gives different results on the penetration 
depth anisotropy on single
crystal and powder MgB$_2$ samples\cite{Cubitt, Cubitt06, 
Eskildsen06}. In general, it is more of a
challenge to determine the absolute value of the penetration depth as 
compared with its temperature
dependence. Although muon spin resonance ($\mu$SR)\cite{Son00} and 
nuclear magnetic resonance (NMR)
methods\cite{Rey97} have often been used to obtain an absolute value 
of the penetration depth, the
application of these resonance techniques to determine the 
penetration depth for an anisotropic
superconductor with a sample consisting of a randomly oriented powder 
has never been attempted until
now.

NMR and electron spin resonance (ESR) have been used previously to
investigate the anisotropy of MgB$_2$.  Two different components of 
the resonance signal have been
identified in the superconductive state in a restricted range of magnetic
field\cite{Simon,Papavassiliou, Lee04} and the anisotropy of the 
upper critical field has been
deduced. Additionally, an attempt was made to determine the 
temperature dependent penetration depth
from the NMR linewidth\cite{Lee04} assuming that MgB$_2$ is 
isotropic, which is clearly not the case.

\begin{figure}[ht]
\vspace{0.1in}
\includegraphics[width=8cm]{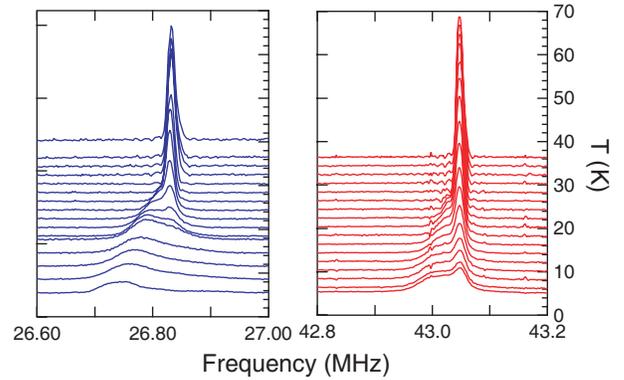}
\caption[$^{11}$B NMR spectra]{$^{11}$B NMR spectra of the central 
transition in magnetic fields of
1.97 T (left) and 3.15 T (right) obtained from a frequency sweep 
described in the text. The $(\pm 3/2 \leftrightarrow \pm 1/2)$
satellites are about 340 kHz away from the central transition and, 
although they are not shown
here, they are shown in Fig.6.}
\label{Fig1}
\end{figure}
Here we report ${}^{11}$B NMR measurements in the temperature range 
from 5 K to 40 K on a
powder sample of MgB$_2$ at two magnetic fields, 1.97 T and 3.15 T. 
We find that on cooling the
spectra acquire a broad asymmetric line below the superconductive 
transition temperature as shown in
Fig.1. The shape of the broad line suggests the expected lineshape 
from an inhomogeneous
field distribution from vortices in their solid state. However, as we 
will see, this interpretation
is too simplistic.  The relative weight of this broad line, compared 
to the narrow normal component,
increases with decreasing temperature. We associate this with the 
temperature and angular dependence
of the upper critical field. From this behavior we obtain the 
upper critical field anisotropy to
be 5.4 at low temperature. We have also simulated the full spectrum 
at 8 K in a field of 1.97 T using
this value for anisotropy and, by comparing with experiment, we have 
obtained the penetration depth
$\lambda = 1,152\pm50$ {\AA}.  Furthermore, we find that the 
penetration depth is isotropic for \mbox{T $ < 10 $ K} even
though the upper critical field and the coherence length are not. 
Our results support the
theoretical claim that there are two different anisotropy parameters 
for upper critical field and
penetration depth\cite{Kogan1,Golubov,Kogan2}.

The polycrystalline MgB$_2$ sample was prepared by solid state 
reaction techniques using a mixture
of magnesium and boron powders. The superconductive transition 
temperature was measured to be 39.5 K
for the onset of diamagnetism in a magnetic field of 1.0 mT and  39 K 
for zero resistance. A sample
of 0.2 gram randomly oriented MgB$_2$ powder was used in our 
experiments. NMR measurements were
carried out in the temperature range between 5 K and 40 K in magnetic 
fields of 1.97 T and 3.15 T in a
superconductive magnet. Broad spectra were obtained by summing 
Fourier transforms of echo
signals for a suite of different frequencies that cover the NMR spectrum.

The spectra displayed in Fig. 1 are the central transition 
$\left(-1/2\leftrightarrow1/2\right)$ of
${}^{11}$B. At high temperature, the sample is metallic in the normal 
state and this spectrum
consists of a single narrow and symmetric line.  As the temperature 
is lowered, a broad and
asymmetric line appears. We associate this with the inhomogeneous 
field distribution from
vortices in the superconductive state in addition to diamagnetic 
screening currents\cite{Brandt}. The
weight of the broad line increases with decreasing temperature, while 
that of the narrow line
decreases.  The two lines coexist to a temperature of 5 K at 3.15 T, 
whereas only the broad line
survives below 17 K at 1.97 T. This can be explained by anisotropy of 
the upper critical field in
MgB$_{2}$.

\begin{figure}[t]
\vspace{0.1in}
\includegraphics[width=8cm]{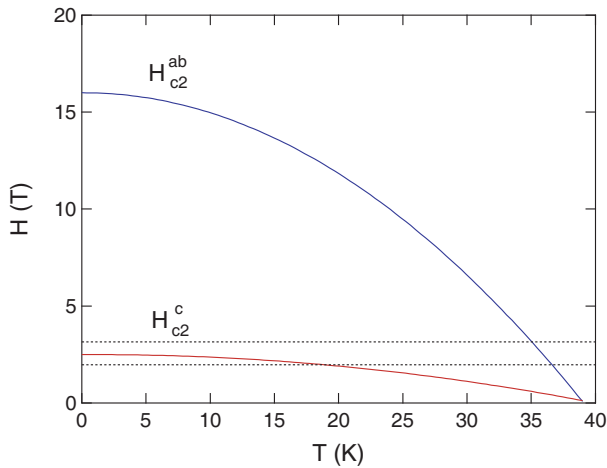}

\caption[H$_{c2}$ of MgB$_2$]{Schematic H$_{c2}$ diagram of MgB$_2$. 
The upper dotted line denotes
3.15 T, and the lower line corresponds to 1.97 T. The value for 
H$_{c2}(0)$ is taken from Bud'ko
and Canfield\cite{Bud'ko}.}
\label{Fig2}
\end{figure}

Due to the temperature dependence and anisotropy of the two gap parameters in
MgB$_2$\cite{Kogan1,Kogan2,Choi}, its upper critical field has a 
temperature and angular dependence:
\begin{eqnarray} 
H_{c2}(\theta,T)=H^{ab}_{c2}(T)/\sqrt{1+(\gamma^2_{H}-1)\cos^2\theta}
\end{eqnarray} where $\theta$ is the angle between the applied 
magnetic field and the $c-$axis of a
crystal. The temperature dependence of H$^c_{c2}$ and H$^{ab}_{c2}$ 
is sketched in
Fig. 2. For temperatures below $T_{c}(H)$, the upper critical
field H$_{c2}$ will be equal to the applied magnetic field for 
crystals oriented at a certain angle
$\theta_{cr}(T)$. The crystals with $\theta$ larger than
$\theta_{cr}(T)$ have their H$_{c2}$ greater than the applied field, 
and are superconductive. Due to
the random distribution of the orientation of the crystals, the 
superconductive fraction in the
sample simply equals cos
$\theta_{cr}(T)$.  As the temperature decreases further, H$^c_{c2}$ 
increases and, if it crosses the
applied magnetic field, the whole sample becomes superconductive. In 
Fig. 1, for H = 1.97 T, the
narrow line disappears below 17 K leaving only the broad line. 
However, in H = 3.15 T the
$c-$axis upper critical field, H$^c_{c2}$, is always smaller than the 
applied field. Therefore, part
of the sample remains in the normal state in this field and 
contributes to the narrow line in the
spectrum, even at the lowest temperatures.

In Fig. 1, the position of the narrow peak is almost temperature 
independent and has a
gaussian shape. Therefore, the contribution of crystals in the normal 
state can be deconvolved from
the composite spectra. The ratio of the remaining area to the whole 
spectrum gives the
superconductive fraction cos $\theta_{cr}(T)$, plotted in Fig. 3. 
Furthermore, with H$^{ab}_{c2}(0)$
taken to be 16 T from Bud'ko and Canfield\cite{Bud'ko}, and the upper 
critical field at $\theta_{cr}$ at 5 K equal to 3.15 T, the external 
applied magnetic field, the upper critical field anisotropy
$\gamma_H$ can be obtained from Eq. 1 and we find this to be 5.4 at 
low temperature. This value is
consistent with
previous
reports\cite{Cubitt,Papavassiliou,Eskildsen,Fletcher,Angst,Lyard,Angst1,Zehetmayer,Kogan1,Golubov,Kogan2}.

Assuming $\gamma_H$ to be temperature independent, H$^{ab}_{c2}$ at 
each temperature point can be
obtained following Eq. 1. The temperature dependence for this 
analysis is plotted in Fig. 4 where it
is compared with results  for H$^{ab}_{c2}$ from other 
groups\cite{Bud'ko,Simon}. The discrepancy
grows with increasing temperature. However, it is now accepted that 
$\gamma_H$ decreases with
increasing temperature\cite{Kogan1,Kogan2}. Therefore, in our 
derivation at high temperatures we have
used a value for $\gamma_H$ that is too large which will produce a 
larger H$^{ab}_{c2}$ and
consequently an overestimate of the transition temperature in a given 
field.   In contrast,
Fig. 4 shows that the critical field curve deduced from our data and 
Eq. 1 is too low.  In fact, it
extrapolates to a zero field transition temperature around 30 K.  The 
principal reason for this
discrepancy is vortex dynamics.
  At high temperatures vortices are in a liquid 
state\cite{Maple,Rey97} and their dynamics on the NMR
time scale average the local fields to zero at the
$^{11}B$ nucleus.  This transfers spectral weight from the broad line to the
narrow line and reduces the apparent superconductive fraction 
obtained from NMR. At low temperatures,
in the vortex solid state our analysis of the superconductive 
fraction is reliable and, as can be
seen in Fig. 4, our results match  H$^{ab}_{c2}(T)$ below 10 K.

\begin{figure}[t]
\vspace{0.1in}
\includegraphics[width=8cm]{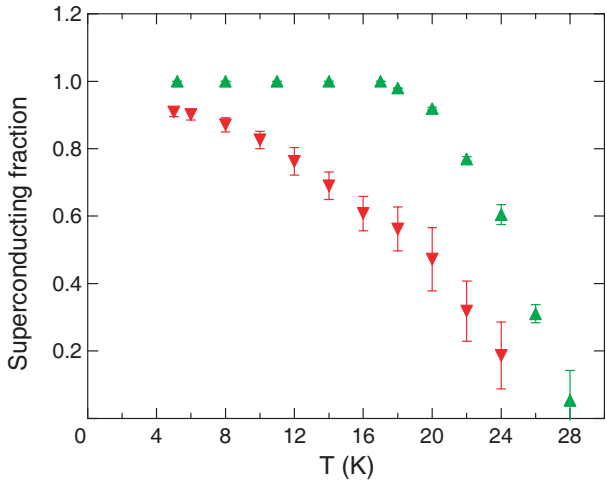}

\caption[Temperature dependence of the fraction of superconductive crystallites
compared to those that are not superconducting.]{Temperature 
dependence of the fraction of
superconductive crystallites in the sample
determined from the composite spectra plotted vs temperature for 1.97 
T (uptriangle) and 3.15 T
(downtriangle).}
\label{Fig3}
\end{figure}

\begin{figure}[ht]
\vspace{0.1in}
\includegraphics[width=8cm]{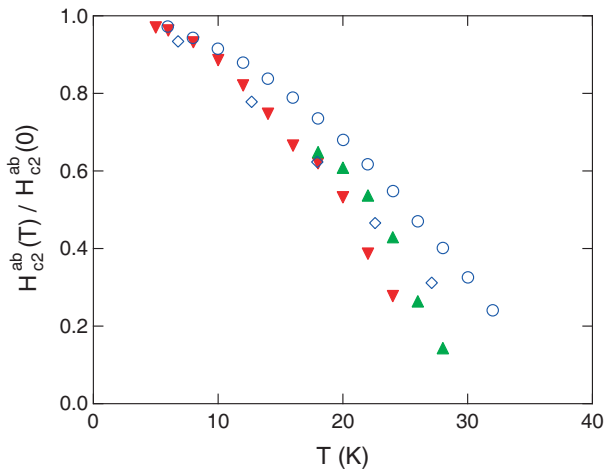}

\caption[Normalized H$^{ab}_{c2}$]{H$^{ab}_{c2}$. The
circles are results from F. Simon {\it et al.}\cite{Simon} (circle) 
and diamonds are from S. L.
Bud'ko and Canfield\cite{Bud'ko}. Analysis of our data using Eq. 1 is 
plotted assuming a constant
$\gamma_H$ = 5.4 for 1.97 T (uptriangle) and 3.15 T (downtriangle). 
We infer that our interpretation
of the NMR signal as a vortex-broadened, inhomogeneous magnetic field 
distribution, is valid only
below 10 K. }
\label{Fig4}
\end{figure}

We have found that the  spin-lattice relaxation rate of the broad 
line is much slower than that of the
narrow line.   This agrees with a previous report\cite{Karayanni}. Additionally, 
we have found that
the rate increases smoothly with increasing frequency within the 
spectrum, rising in the high
frequency tail of the central transition.  Owing to the inherent 
inhomogeneity of the field
distribution, which we will discuss later, it is not possible to 
deconvolute spin-lattice relaxation
signals to search for electronic excitations in different parts of the 
vortex structure as has been
reported\cite{mit01} for YBCO. However, our spin-lattice relaxation 
results serve as a guide to help
us avoid selective saturation, particularly at low frequencies where 
the rate is small, allowing us to
obtain a faithful representation of the spectrum.

The absolute value of the penetration depth $\lambda$ is a key 
parameter for characterization of
superconductivity and yet it is difficult to measure accurately. 
Using a tunnel diode oscillator
technique Fletcher {\it et al. }\cite{Fletcher} found a penetration 
depth of MgB$_2$ between 800 and
$1,200$ \AA. Finnemore {\it et al.}\cite{Finnemore} determined that
$\lambda_{ab}$ was $1,400$ {\AA} from transport measurements.  Using 
ESR, Simon {\it et
al.}\cite{Simon} reported a value of the penetration depth between 
$1,100$ {\AA} and $1,400$ {\AA}. From
analysis of the second moment of the measured NMR linewidth,  Lee 
{\it et al.} calculated the
penetration depth to be $2,100$
\AA. But, as we mentioned earlier, the resonance methods cannot 
obtain a reliable measure of the
penetration depth, if it is assumed in their interpretation that the 
superconductor is isotropic.
Here we determine the penetration depth by comparison of our measured 
spectrum with a simulation of
the local fields in the mixed state for an anisotropic random powder 
at 8 K in a magnetic field of
1.97 T using the penetration depth as a variational parameter.

The NMR spectrum is a local magnetic field map. At low temperature, 
the vortices are in the solid
state and contribute to the associated field distribution of the NMR 
spectrum. The
field distribution of the mixed state can be calculated by solving 
the Ginzburg-Landau (GL)
equation. For this purpose we adopt Brandt's algorithm\cite{Brandt} an
iterative, quickly converging method. The solution gives the current and field
distribution from the vortex lattice and the diamagnetic fields from 
screening currents in the
superconducting state. The required inputs are the external field, 
coherence length $\xi$, and
penetration depth $\lambda$. We calculate the coherence length, 
$\xi_{ab} = 108$ {\AA} from the upper
critical field\cite{Bud'ko} and we take its anisotropy from Eq. 1,

\begin{eqnarray}
\xi(\theta)=\xi_{ab}/\sqrt{1+(\gamma_{H}^2-1)cos^2\theta}
\end{eqnarray}
With these inputs, the field distribution for a crystal at a specific 
angle is generated by
Brandt's algorithm\cite{Brandt} including the central transition and 
its quadrupolar
satellites. We convolute the spectrum with a broadening 
function, exp$-2(H/\delta)^2$, which will also include the effect of the finite width of the 
NMR line in the normal state. In a powder sample, which we assume to 
be composed of single crystal ellipsoids of revolution, we must consider the
shifts of magnetization owing to demagnetization according to the 
shape and orientation distribution
of the grains\cite{Drain}. For simplicity we characterize this 
distribution by an average
demagnetization factor, D. This assumption would be precise if the 
grain shape distribution is
uncorrelated with the crystal structure. The demagnetization effect 
gives a relative shift of the
magnetization which itself depends on the orientation of the grains 
since the diamagnetic moment from
screening currents is strongly angular dependent.  Simulations of 
spectra at three different, but
representative, angles are presented in Fig. 5. The spectrum for the 
whole sample is then obtained as
the integral of 91 spectra with orientation uniformly distributed between 0 and
$\pi/2$, weighted by a factor sin $\theta$ appropriate for a random 
distribution of
grain orientations.

\begin{figure}[ht]
\vspace{0.1in}
\includegraphics[width=8cm]{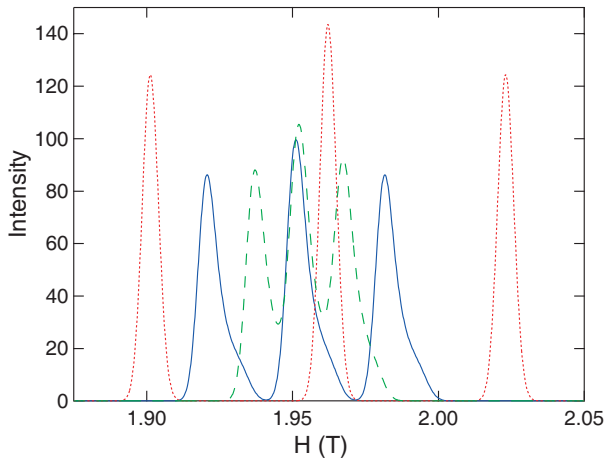}

\caption[Simulated spectra for different orientations]{Simulated 
spectra for crystals at different
orientations. The solid, dashed and dotted curves are the spectra of 
crystals with $c-$axis at
$\pi/2$, $\pi/4$ and 0 angles to the applied field, respectively.  A 
demagnetization
factor, D = 1/3,  and a gaussian broadening parameter, $\delta = 
5.2$ mT,  were chosen for these
spectra.}
\label{Fig5}
\end{figure}

There are three
variational parameters,
$\lambda$, D, and $\delta$.  We then carry out a
$\chi^2$ minimization of the difference between the simulated 
spectrum and the experimental one,
taking their areas to be equal. The simulated spectrum is shown in 
Fig. 6 together with the
experimental spectrum. The numerical results provide an excellent 
representation of the complex
measured spectrum with values for the variational parameters for the 
penetration depth
$\lambda$ =
$1,152\pm50$ \AA, average demagnetization factor, D = $0.31\pm0.01$ 
and the gaussian broadening,
$\delta = 5.2$ mT, which is larger than, but of the same order as, the normal
state linewidth, 2 mT.  The  quoted accuracy is
statistical.  This value for D is rather close to that  anticipated 
for a spherical geometry,
D$_{sphere} = 1/3$, and it is reasonable to expect this value for the 
average demagnetization factor
for a large ensemble of grains.  Earlier reports for the value of the 
penetration
depth\cite{Simon,Fletcher,Heon,Finnemore}  are similar to ours 
although our accuracy is higher.  Our
simulation and its comparison with experiment, as represented in Fig. 
6, is the most precise such
comparison obtained by resonance methods and it is the first time 
that such a simulation has  been
attempted for a strongly anisotropic supercondcutor. We emphasize 
that previous work has generally
focused on moments of the measured spectrum, often restricting 
attention to the second moment.
For an anisotropic superconductor the angular dependence of the first 
moment of the distribution, as
can be seen in Fig. 5, must be correctly handled since it contributes 
significantly to the
overall lineshape.  Earlier work on other superconductors analyzing the field
distribution in the mixed state has been directed at the moments of 
the distribution, so we have
calculated the first three moments for an anisotropic superconductor 
with randomly oriented grains,
as a function of the penetration depth, restricted to the case of
$\gamma_{H} = 5.4$ and $\gamma_{\lambda} = 1$.

\begin{figure}[ht]
\vspace{0.1in}
\includegraphics[width=8cm]{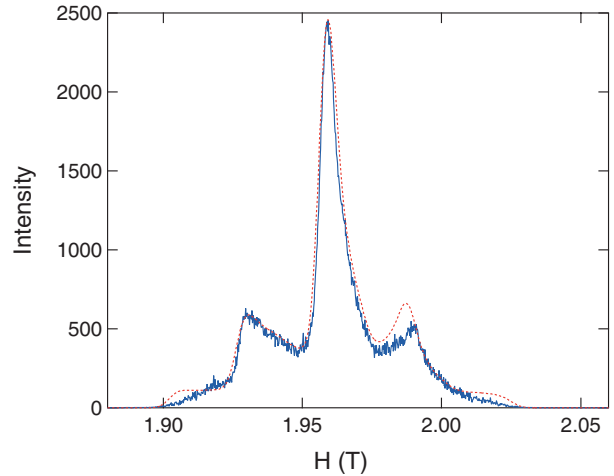}

\caption[The simulated spectrum of 8 K at 1.97 T.]{The spectrum at a 
temperature of 8 K and a
magnetic field of 1.97 T. The blue solid line is the experimental 
spectrum. The dotted line is the
simulation described in the text.}
\label{Fig6}
\end{figure}

\begin{figure}[t]
\includegraphics[width=8cm]{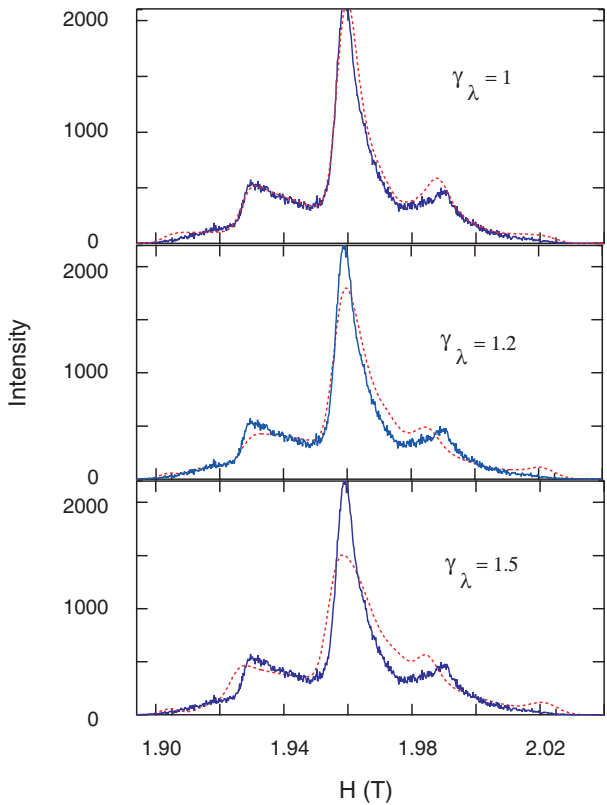}

\caption[Simulation with different $\gamma_{\lambda}$.]{Simulation 
with different
$\gamma_{\lambda}$ plotted together with experimental spectrum at a 
temperature of 8 K and a
magnetic field of 1.97 T. The dashed lines are simulations. It is 
clear that the best comparison
between experiment and simulation holds for $\gamma_{\lambda} \approx 1$.}
\label{Fig7}
\end{figure}

The second moment of the magnetic field distribution of a spectrum 
from a vortex lattice
can be related\cite{Brandt} to its penetration depth $\lambda$ for 
low magnetic fields compared to
H$_{c2}$ by the Pincus' formula\cite{Brandt,Rey97} where the second
moment varies  as the inverse fourth power of the
penetration depth, $\left\langle{B^2}\right\rangle=
(0.0609 \phi_{0})^{2}/\lambda^4$. In the present case the simulated 
spectrum is the superposition of
spectra with
anisotropic coherence lengths and upper critical fields. 
Nonetheless, we find the 1st, 2nd and 3rd
moments of the spectrum can be similarly related to inverse, even 
powers of the penetration depth in
the following elegant way:

\begin{eqnarray}
\left\langle{B}\right\rangle=-(1-D)\cdot{A_1}/\lambda^2
\end{eqnarray}
\begin{eqnarray}
\left\langle{B^2}\right\rangle= \delta^2/4 + A_2/\lambda^4
\end{eqnarray}
\begin{eqnarray}
\left\langle{B^3}\right\rangle=A_3/\lambda^6,
\end{eqnarray}
where A$_{1}$, A$_{2}$, A$_{3}$ are numerical constants. The gaussian 
broadening factor is $\delta$
and D is the demagnetization factor. We find A$_1$ = 1.415 $\times$ 
10$^4$ T \AA$^2$,
\mbox{A$_2$ = 2.621
$\times$ 10$^7$ T$^2$
\AA$^4$}, A$_3$ = 1.524 $\times$ 10$^{11}$ T$^3$ \AA$^6$. However we 
caution that anisotropy and field
dependence of the local field distributions mean that these numerical 
constants hold only in a limited
range which we have explored for  MgB$_2$ with $\gamma_{H} = 5.4$, 
$\gamma_{\lambda} = 1$ and H =
1.97 T.

We have also investigated the effect of the penetration depth 
anisotropy $\gamma_{\lambda}$ at low
temperature. In actuality, the vortex structure for arbitrary
angle $\theta$ is found as the solution to the anisotropic GL 
equations\cite{Oliveira}. However, as a
reasonable approximation for an almost isotropic penetration depth, 
we introduce another variational
parameter,
$\gamma_{\lambda}$, and continue to adopt the solution of the
isotropic GL equation for each crystallite.  The anisotropy of 
H$_{c1}$ is the inverse of H$_{c2}$,
therefore the penetration depth has the inverse angular dependence of 
the coherence length,

\begin{eqnarray}
\lambda(\theta)=\lambda_{ab}\sqrt{1+(\gamma_{\lambda}^2-1)cos^2\theta}
\end{eqnarray}

With the same approach as before we generate the spectrum for the 
powder sample with
$\gamma_{\lambda}$ larger than one. The central transition is found 
to decrease with increase of
$\gamma_{\lambda}$ and the spectra become more asymmetric as Fig. 7
illustrates. This suggests that $\gamma_{\lambda}$ at low 
temperatures should be close to one.
Magnetization measurements show that $\gamma_{\lambda}$ is around 1.7 
between 20 and 27
K\cite{Heon}. SANS experiments on a powder MgB$_2$ 
sample\cite{Cubitt06} give an upper limit of
$\gamma_{\lambda}$ to be around 1.5 and essentially magnetic field 
independent. Our result is
consistent with these values. However, SANS measurements on a single 
crystal\cite{Cubitt,Eskildsen06}
indicate that $\gamma_{\lambda}$ is close to one at T = 2 K and at 
low filed, H $<$ 0.5 T, and
that it increases with external field reaching $\approx$ 3.5 in a 
field of 0.8 T. The lower
$\gamma_{\lambda}$ in the powder sample is believed to be caused by a 
limiting crystallite size
effect\cite{Cubitt06}. Further work will be required to elucidate 
this phenomenon.

In conclusion, we measure the $^{11}$B NMR spectra of a random powder 
sample of MgB$_2$ in magnetic
fields of \mbox{1.97 T} and 3.15 T. The evolution of the spectra through the 
temperature range can be
explained by the anisotropy of the upper critical field $\gamma_{H}$, 
which is determined to be 5.4
at low temperature. We find from our simulation that the
penetration depth carries a different anisotropy from the upper 
critical field and that at low
temperatures it is almost isotropic similar to that reported from 
SANS\cite{Cubitt06} for a powder
sample.  The value of the penetration depth that we have obtained for 
MgB$_2$ is
$1,152\pm50$ {\AA} at 8 K in a magnetic field of 1.97 T. From our 
numerical studies we have found
simple expressions for the penetration depth dependence of the 
moments of the field distribution in a
random powder of an anisotropic superconductor.  

This work was supported by the DOE:
DE-FG02-05ER46248. Two of us (M. Lee and B. K. Cho) acknowledge financial support 
from the Korean Research Foundation through, respectively, Grant 
2003-015-C00161 and ABRL program at Ehwa Woman University.

\end{document}